\documentclass[10pt,twocolumn,letterpaper]{article}

\usepackage{cvpr}              

\usepackage{amsmath}
\usepackage{bbm}
\usepackage{color}
\usepackage{wrapfig,lipsum,booktabs}
\usepackage{graphicx}
\usepackage{float}
\usepackage{multirow}
\usepackage{makecell}
\usepackage{array}
\usepackage{diagbox}
\usepackage{arydshln}
\usepackage{stfloats}

\usepackage{algorithm}
\usepackage{algorithmicx}
\usepackage[noend]{algpseudocode}

\usepackage[dvipsnames]{xcolor}

\title{SGNet: Folding Symmetrical Protein Complex with Deep Learning}

\newcommand{\authorskip}{\hspace{6mm}}

\author{Zhaoqun Li
\authorskip Jingcheng Yu
\authorskip Qiwei Ye
\\[5mm]
Beijing Academy of Artificial Intelligence}

\begin{document}
\maketitle
\begin{abstract}
Deep learning has made significant progress in protein structure prediction, advancing the development of computational biology. 
However, despite the high accuracy achieved in predicting single-chain structures, a significant number of large homo-oligomeric assemblies exhibit internal symmetry, posing a major challenge in structure determination. 
The performances of existing deep learning methods are limited since the symmetrical protein assembly usually has a long sequence, making structural computation infeasible. 
In addition, multiple identical subunits in symmetrical protein complex cause the issue of supervision ambiguity in label assignment, requiring a consistent structure modeling for the training.
To tackle these problems, we propose a protein folding framework called SGNet to model protein-protein interactions in symmetrical assemblies. 
SGNet conducts feature extraction on a single subunit and generates the whole assembly using our proposed symmetry module, which largely mitigates computational problems caused by sequence length.
Thanks to the elaborate design of modeling symmetry consistently, we can model all global symmetry types in quaternary protein structure prediction. 
Extensive experimental results on a benchmark of symmetrical protein complexes further demonstrate the effectiveness of our method.

\end{abstract}

\section{Introduction}

Protein structure prediction is a crucial step in understanding complex protein functionality. 
The way in which proteins interact with each other provides insight into how biological processes occur at the molecular level~\cite{bryant2022improved, zeng2018complexcontact, jin2022Iterative}. 
Recent deep learning techniques~\cite{stark2022equibind, baek2021accurate, huang20223dlinker, gao2022af2complex, tsaban2022harnessing, Richard2021alphafoldmultimer, ganea2021independent, krapp2023pesto} have been widely applied in this domain, leading to the development of novel drug discovery, protein engineering, protein-ligand binding, and more. 
AlphaFold~\cite{jumper2021highly,Richard2021alphafoldmultimer} has shown promising performance in predicting complex structures; however, predicting the folding of large homo-oligomeric protein complexes remains a major challenge.

A significant category of protein-protein interactions is formed by homo-oligomers with specific symmetry~\cite{gaber2021modeling, li2022uni}, especially for large protein complexes. 
For example, virus capsids are typically composed of many identical proteins arranged with icosahedral symmetry. Therefore, modeling and predicting symmetrical protein complexes are critical for understanding the fundamental mechanisms involved in protein folding, shedding light on general multi-chain protein structure prediction. 
Traditional computational methods typically use symmetry docking algorithms to tackle this problem, such as HSYDOCK~\cite{yan2018hsymdock} and SymDock2~\cite{burman2019flexible}. 
Unfortunately, the docking process typically suffers from both low accuracy and long latency. Certain recent studies~\cite{mirdita2022colabfold, bryant2022improved, gao2022af2complex, ghani2022improved, bryant2022predicting, chen2022improved} have been able to predict multimeric interfaces and attain promising performance. 
To overcome the issue of label assignment permutation symmetry in homomeric components, AlphaFold-Multimer (AFM)~\cite{Richard2021alphafoldmultimer} proposes multi-chain permutation alignment that greedily searches for a good permutation during training.
AFM achieves high accuracy in predicting symmetrical structures, yet it faces challenges in terms of residue length restrictions and the greedy search process lacks stability during training of cubic symmetrical complexes. 
Leveraging the symmetry group property, UF-Symmetry~\cite{li2022uni} designs a training regime to hasten symmetry complex folding and assemble the entire protein in an end-to-end manner.

Despite numerous efforts devoted to the field, predicting the structure of symmetrical proteins using deep learning remains a challenging problem. The task primarily encounters two major obstacles. 
Firstly, the sequence of a symmetrical protein assembly is typically very long as it consists of multiple subunits. Popular methods like AFM have cubic complexity and suffer from the notorious sequence length issue, which may cause computational infeasibility. Secondly, another difficulty arises from the symmetrical structure, which brings a label assignment issue in supervised learning. As the subunits are identical sequences, a fully sequence-to-structure model would encounter supervision ambiguity. Therefore, a proper training scheme that considers the symmetry characteristics is also desirable.

To address this gap, we propose a methodology for modeling the symmetrical protein quaternary structure that leverages protein-protein interface properties and underlying inter-chain relations. 
By enumerating and analyzing all structural topologies in symmetrical protein assemblies, we unify the symmetry relation representation using relative position maps and design a symmetry generator that can duplicate subunits to reconstruct the whole assembly. 
All global symmetry types in proteins can be modeled in this approach. 
Additionally, based on the structure modeling, we propose Symmetry Generator Network (SGNet), which is a neural framework specially designed for symmetrical protein folding. 
In the framework, the sequence feature extractor is based on evoformers and leverages single-chain feature embedding and geometric information extracted from AFM.
The symmetry module takes the pretrained single-chain feature as input to construct the whole symmetrical protein assembly.
In this way, we can largely alleviate the computational problems caused by sequence length. 
And during the training stage, the learning objectives are modeled consistently to avoid ambiguity issues.
Experimental results demonstrate that our framework can model protein symmetry and achieve better performance than the baseline method, AlphaFold-Multimer.
The contributions are summarized as follows:

\begin{itemize}
    \item We develop folding algorithms that specifically model global symmetry in protein complex structure, including Cyclic, Dihedral, Tetrahedral, Octahedral and Icosahedral symmetries.
    \item We implement a deep model named SGNet for symmetrical protein complex structure prediction leveraging inter-chain interfaces, achieving competitive performance on quaternary structure prediction.
\end{itemize}

\section{Related work}
Most protein complexes tend to exhibit symmetry,
with multiple identical subunits forming the structure and interacting with neighboring subunits in the same manner. 
Predicting the whole protein complex structure is challenging since the protein is usually huge. 
Traditional computational methods typically use symmetry docking algorithms to tackle this problem~\cite{ritchie2016spherical,park2019galaxytongdock,burman2019flexible}. 
HSYDOCK~\cite{yan2018hsymdock} presents a computational tool designed to predict the structure of protein homo-oligomers that have specific types of symmetry, namely cyclic or dihedral. These types of symmetries are common in homo-oligomeric proteins, which are complexes made up of several identical subunits.
SymDock2~\cite{burman2019flexible} focuses on a computational method or framework for assembling and refining the three-dimensional structures of symmetrical homomeric complexes, where multiple identical protein subunits assemble in a specific symmetric pattern.
Regrettably, the docking process often experiences both limited precision and extended delays.

Certain recent studies~\cite{mirdita2022colabfold, Jeppesen2023Accurate, bryant2022improved, gao2022af2complex, ghani2022improved, bryant2022predicting, chen2022improved, watson2023novo} have been able to predict multimeric interfaces and attain promising performance. 
AlphaFold~\cite{jumper2021highly,Richard2021alphafoldmultimer} has demonstrated impressive capabilities in predicting the structures of both monomeric and multimeric proteins; however, it faces constraints related to the count of protein chains and the number of amino acid residues it can process.
In~\cite{Jeppesen2023Accurate}, authors present a method that combines AlphaFold's capabilities with all-atom symmetric docking simulations to predict the structure of complex symmetrical protein assemblies.
This approach allows for the energetic optimization of models and aids in the study of intermolecular interactions within symmetrical assemblies.
RFdiffusion~\cite{watson2023novo} introduces a deep-learning framework for designing proteins, addressing a broad spectrum of design challenges, including the creation of new binders and symmetric structures. The RoseTTAFold structure leads to a generative model for protein backbones with high performance in various design applications. However, the framework is not designed for symmetric structure prediction.

\section{Preliminaries}


\subsection{Symmetry group}
The symmetry group of a geometric object is the group of all transformations under which the object is invariant.
Oligomeric assemblies can exhibit two categories of symmetries: point group symmetry or helical symmetry~\cite{david2000Structure}.
The paper narrows its focus specifically on finite point groups including five types:
Cyclic groups (C$n$), Dihedral groups (D$n$),
Tetrahedral group (T), Octahedral group (O) and Icosahedral group (I).



In the first two symmetry symbols, $n$ means the symmetry group has $n$-fold cyclic axis. 
Among these symmetry groups, T, O and I symmetries are cubic symmetries which are more challenging to dissolve.


\subsection{Backbone frames}
We follow~\cite{jumper2021highly} to represent one backbone frame via a tuple $T:(R,t) \in \mathrm{SE(3)}$. 
In the tuple, $R$ is the rotation matrix, $t$ is the translation vector which is also the position of $C^\alpha$ by definition. 
The backbone of a protein chain with length $N$ can be represented as a sequence $\{T_i\}_{i\leq N}$. 
Each backbone frame also represents a 3D Euclidean transform from the local frame to a global reference frame:
\begin{equation}
    x_{global}=T\circ x_{local}=R x_{local}+t  
\end{equation}
The composition of two Euclidean transforms is denoted as
\begin{equation}
    T_{comp} = T_1\circ T_2=(R_1R_2, R_1t_2+t_1)
\end{equation}
The relative position of one point $x \in \mathbb{R}^3$ with respect to a coordinate system $T$ is the local coordinate of $x$ in $T$:
\begin{equation}
    x_{local} = T^{-1}\circ x_{global} = R^T(x_{global}-t) 
\end{equation}
For convenience, we use $\mathcal{T}$ to represent all backbone frames in a chain.

\section{Structural symmetry modeling}
\label{sec_modeling}
In this section, we describe the symmetry modeling from a structural topology view and explain how to formulate the learning objectives.
Since residues are basic units of a protein chain, the symmetry structure modeling description solely involves the backbone pose, i.e. $\mathcal{T}$, and omits the side-chains. 




\subsection{Protein symmetry}
For a symmetrical protein complex, its amino acid sequence is composed of multiple identical subsequences corresponding to identical polypeptide chains. The smallest subsequence is called the asymmetric unit (ASU), and we can also regard it as one protein chain. For example, the protein \textit{1nw4} 
has D3 symmetry and its sequence is $\{A_1A_2A_3A_4A_5A_6\}$, in which each chain $A_i$ is identical to another one and the number of duplicates is 6. 
We can reconstruct the whole assembly from the ASU if we know its positional relation with its neighboring chains.

\textbf{Protein contact interface.} 
From a perspective of structural topology~\cite{ahnert2015principles}, 
there are two types of interfaces between subunits in a symmetry assembly: heterologous interfaces and isologous interfaces.
In the topology, ASUs are viewed as graph nodes, and the contact interfaces are the graph edges.
Then we can enumerate all quaternary structure topologies in symmetrical protein assembly (see Appendix~\ref{sec_topology}).
There are various cases in structural topologies, even for one specific symmetry. The intricate cases present a formidable challenge in the field of symmetry modeling when attempting to achieve a satisfactory outcome.


\subsection{Relative position map}
\label{sec_rel_pos}

\begin{figure}
    \centering
    \includegraphics[width=0.47\textwidth]{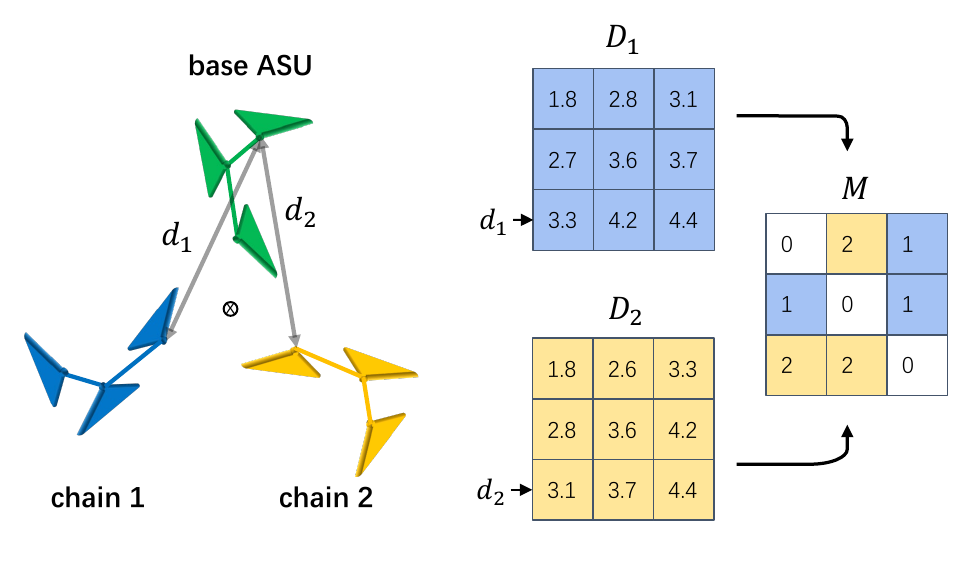}
    \caption{An illustration example of chain index mask generation. $\otimes$ is the symmetry axis.} 
    \label{fig_rel}
\end{figure}

The symmetric relationship in 3D space can be described by relative position map.
Let $\mathcal{T}^a,\mathcal{T}^b$ be the backbone frames of two protein chains in a symmetrical assembly and assume the chain lengths are $N_a, N_b$.
Denote $t^{a} \in \mathbb{R}^{N_a\times 3}$, $t^{b} \in \mathbb{R}^{N_b\times 3}$ as the translation vectors. 
The relative position map between the two protein chains is defined as the $C^{\alpha} $ local coordinate map:
\begin{equation}
\begin{aligned}
    P^r_{a,b} &\in \mathbb{R}^{N_a\times N_b\times 3} \\
    P^r_{a,b}(i,j)&=\mathcal{T}_i^{a^{-1}}\circ t^b_j
\end{aligned}
\end{equation}
that is, the pixel $(i,j)$ in the relative position map stores the local coordinate of $j$-th $C^{\alpha}$ in chain $b$ with respect to $i$-the backbone frame $\mathcal{T}^a_i$.
An important property of the representation is that $P^r$ is \textit{SE(3)-invariant}.
This can be obtained by applying an arbitrary transformation $T_g \in \mathrm{SE(3)}$ to both chains:
\begin{equation}
\begin{aligned}
    (T_g\circ\mathcal{T}_i^{a})^{-1}\circ (T_g\circ t^b_j) &=\mathcal{T}_i^{a^{-1}}\circ T_g^{-1}\circ T_g\circ t^b_j \\
    &=\mathcal{T}_i^{a^{-1}}\circ t^b_j
\end{aligned}
\end{equation}


The isologous/heterologous interfaces can be determined by relative position map and then we can calculate the symmetry axes.
We implement empirical algorithms for the calculation, which are described in detail in Appendix (Alg.~\ref{alg_iso2axis} and Alg.~\ref{alg_hetero2axis}).

\begin{algorithm}[t]
\caption{D2 symmetry generator}
\label{alg_d_sym1}
    \begin{algorithmic}[1]
    \renewcommand{\algorithmicrequire}{\textbf{Input:}}
    \renewcommand{\algorithmicensure}{\textbf{Output:}}
    \Require Backbone frames $\mathcal{T}$, Confidence score map $C$, Nearest position map $P^{nst}$, Chain index map $M$, Number of selection $K$, Number of replicates $n$.
    \Statex
    \Comment{$P^{nst} \in \mathbb{R}^{N\times N}$, $M \in \mathbb{R}^{7\times N\times N}$, $C \in \mathbb{R}^{7\times N\times N}$} 
    \Ensure Atom positions of backbone $p_{all}$.
    \Statex
    \Comment{$p_{all} \in \mathbb{R}^{n\times N\times 3\times 3}$}

    \renewcommand{\algorithmicrequire}{\textcolor[RGB]{50,140,10}{def}}
    \Require D2SymmetryGenerator($\mathcal{T}$, $P^{nst}$, $M$, $C$, $K$, $n$) 

    \State $p_{bb} := \text{backbone atom positions}$ \Comment{$p_{bb} \in \mathbb{R}^{N\times 3\times3 }$}    
    \State $p_0 := C^{\alpha} \text{ position}$ \Comment{$p_0 \in \mathbb{R}^{N\times 3 }$}
    \State $S = \mathrm{GetPointIndexSet}(M, C, K)$ \Comment{$S \in \mathbb{R}^{7\times K\times 2}$}
    
    \State $l_1 = \mathrm{SymmetryAxisIsologous}(\mathcal{T}, P^{nst}, S[0])$ 
    \State $l_2 = \mathrm{SymmetryAxisIsologous}(\mathcal{T}, P^{nst}, S[1])$ 
    \State $l_3 = \mathrm{SymmetryAxisIsologous}(\mathcal{T}, P^{nst}, S[2])$ 
    \State $p_{1} = \mathrm{ApplySymmetryOp}(p_0, l_1, 2)$ 
    \State $p_{2} = \mathrm{ApplySymmetryOp}(p_0, l_2, 2)$ 
    \State $p_{3} = \mathrm{ApplySymmetryOp}(p_0, l_3, 2)$ 
    \State $l_1, l_2, l_3 = \mathrm{SymmetryAxisIsosceles}(p_0, p_1, p_2, p_3)$ 
    \State $p_{1} = \mathrm{ApplySymmetryOp}(p_{bb}, l_1, 2)$ 
    \State $p_{2} = \mathrm{ApplySymmetryOp}(p_{bb}, l_2, 2)$ 
    \State $p_{3} = \mathrm{ApplySymmetryOp}(p_{bb}, l_3, 2)$ 
    \State $p_{all} = \mathrm{concat}(p_{bb}, p_1, p_2, p_3)$
    
    \State \Return $p_{all}$
    \end{algorithmic}
\end{algorithm}

\subsection{Nearest position map and chain index mask}
\label{sec_nrel_pos}

Our goal is to reconstruct the whole symmetrical assembly using relative position maps of ASU.
However, 
learning $P^r$ is not trivial considering the quotient space property. 
The training should involve permutation invariance in label assignment and in the inference the model should generate one determined structure.
For example, in Fig.~\ref{fig_rel} the sample has C3 symmetry.
The green chain has two equivalent neighbour chains, corresponding two relative position maps $P^r_1, P^r_2$.


Since all subunits are equivalent in the whole assembly, 
without loss of generality, we select one subunit as object of study and call it \textit{base ASU}. 
Under the the assumption that the base ASU only interacts with the nearest residues of neighbor chains, we propose to learn nearest position map:
\begin{equation}
\begin{aligned}
    M_{a}(i,j) &= \mathop{\arg\min}_{k\in \mathcal{N}} D_{a,k}(i,j) \\
    P^{nst}(i,j) &= P^r_{a,M_{a}(i,j)}(i,j) \\
\end{aligned}
\end{equation}
where $D$ is the inter-chain distance matrix defined by pairwise distance of $C^{\alpha}$, $\mathcal{N}$ is the index set of neighbour chains of the base ASU $a$. 
$M$ indicates the chain index map, as shown in Fig.~\ref{fig_rel}.
The neighbour relation is the same as that defined in structural topology.
Note that $P^{nst}$ has no subscript of $a$ because it is independent of the choice of base ASU.
$P^{nst}$ is a symmetric function with $P^r$ as its variables ($P^r_1$ and $P^r_2$ for the sample in Fig.~\ref{fig_rel}) and thus is \emph{invariant} under label assignment.
It also inherits the SE(3)-invariant property from $P^r$.
In addition to $P^{nst}$, we also need to distinguish chain index map $M$ in the reconstruction.
The learning of $M$ is similar to an image segmentation task,
where each pixel is assigned to an integer.

\subsection{Symmetry operation and symmetry generator}
\label{sec_sym_op}

Symmetry operation preserves all the relevant structure of a geometric object.
In protein, it means the whole assembly is unchanged after applying one specific rotational transformation. 
We denote the rotational transformation as $R^{l}_{\theta}: \mathbb{R}^3 \rightarrow \mathbb{R}^3$,
where $l=[l_s, l_e] \in \mathbb{R}^{2\times 3}$ is the rotation axis and $\theta$ is the rotation angle.
Note that a rotation axis is a line without direction, we only use two points on it to represent it to ease the algorithm description.
Using the elements in symmetry group, each symmetry type can connect to a function $G$ that generates the whole assembly by duplicating the ASU:
\begin{equation}
\label{eq_symgenerator}
\begin{aligned}
    G: \mathbb{R}^{N\times 3} \times \mathbb{R}^{K\times 2\times 3} &\rightarrow \mathbb{R}^{N_{all}\times 3} \\
    G(x, \{l_i\}) &= x_{all}
\end{aligned}
\end{equation}
In Eq.~\ref{eq_symgenerator}, $N_{all}=n\times N$ and $n$ is the number of replicates,
$K$ is the number of rotation axes in the symmetry assembly.
We name this function \emph{symmetry generator}.
In our method, we implement these symmetry generators and force the functions to output symmetrical protein structure.
For instance, in C$n$ symmetry there is only one $n$-fold symmetry axis $l$, and we have:
\begin{equation}
\begin{aligned}
G_{\mathrm{C}n}(x, \{l\}) = \oplus\left(R_{\theta_1}^l(x),R_{\theta_2}^l(x), ...,R_{\theta_n}^l(x)\right)
\end{aligned}
\end{equation}
where $\theta_i=\frac{2\pi i}{n}, i=1,2,...,n$ and $\oplus$ is concatenation operation.

As to D symmetry, the situation becomes more complicated.
For $n=3$, three equivalent isologous interfaces need to be considered, corresponding three two-fold symmetry axes $l_1, l_2, l_3$.
In reality, $l_1,l_2,l_3$ should intersect at one point while the predicted symmetry axes are not always consistent.
We design an algorithm described as pseudo code in Alg.~\ref{alg_d_sym1}.
In the algorithm,
we leverage pairwise distances of four chains to build a isosceles tetrahedron. 
Then the ASUs are placed ``appropriately'' on the vertices of the isosceles tetrahedron. 
Details of Alg.~\ref{alg_d_sym1} description and T, O, I symmetry generator implementation are presented in Appendix~\ref{sec_sym_generator}.

\begin{figure*}[ht]
  \centering
  \includegraphics[width=\textwidth]{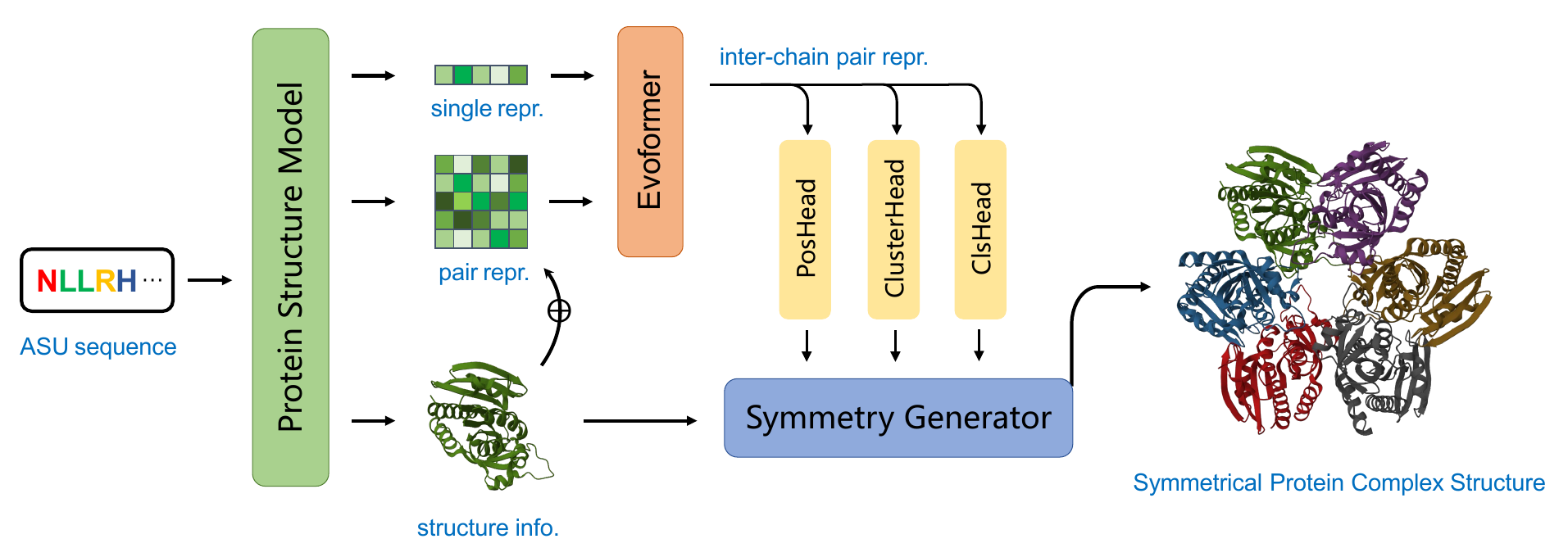}
  \caption{The pipeline of SGNet follows a two-phase, end-to-end process for predicting structure. Initially, the ASU structure is generated using both single and pair representation from a pretrained ASU structure module. Then, the symmetry module expands the ASU to the whole assembly through protein-protein interface information.} 
\label{fig_pipeline}
\end{figure*}

\section{Framework}
This section depicts the overall network pipeline and our proposed symmetry module for predicting the protein structure.
In the following, we denote the number of residues in the input sequence by $N$.

\subsection{Overview}
The overall pipeline is shown in Fig.~\ref{fig_pipeline}. 
There are mainly two parts in our framework,
the ASU structure prediction module and the proposed symmetry module. 
From the protein structure model, which is pretrained in protein folding prediction task,
we can obtain ASU sequence features and its structure:
$s, z, \mathcal{T}, \mathcal{T}_{inter}$.
Here, $s \in \mathbb{R}^{N\times C_s}$ is the single representation, $z \in \mathbb{R}^{N\times N\times C_z}$ is the intra-chain pair representation, 
$\mathcal{T}, \mathcal{T}_{inter}$ are the backbone frames of ASU and its neighbour chains. 
The symmetry module takes ASU structure and inter-chain pair representation as input, generating the whole assembly:
\begin{equation}
   x_{all}= \mathrm{SymmetryModule}(s,z, \mathcal{T}, \mathcal{T}_{inter})
\end{equation}
where $x_{all}$ represents all atom positions in backbones of the symmetrical protein assembly.

\subsection{Pretrain feature embedding}
Before input to the symmetry module, we explicitly extract geometric information from the pretrained protein structure model.
The original pair representation $z$ only have intra-chain information. 
To enhance the inter-chain feature representation, we regard the distance matrix among neighbour chains as a priori knowledge and add it to $z$:
\begin{equation}
\begin{aligned}
z_{input} &= z + \mathrm{Linear}(d_{\mathcal{T}, \mathcal{T}_{inter}})
\end{aligned}
\end{equation}
where $d$ represents inter-chain distance matrix, which is calculated using either 1 or 2 neighbour chains depending on whether interface is isologous or heterologous.

\subsection{Symmetry module}
The objective of the symmetry module is to reconstruct the whole assembly by using protein-protein interfaces and symmetry relations. As discussed in Sec. \ref{sec_rel_pos}, our core idea lies in predicting nearest relative position maps, which provide sufficient symmetry information. To achieve this, in the symmetry module, we add an evoformer stack~\citep{jumper2021highly} to further extract inter-chain features for multimeric interface prediction:
\begin{equation}
\begin{aligned}
z_{inter} &= \mathrm{Evoformer}(s, z_{input})
\end{aligned}
\end{equation}

The resulted feature $z_{inter} \in \mathbb{R}^{N\times N\times C_z}$ is inter-chain pair representation.
After, as shown in Fig.~\ref{fig_pipeline}, three lightweight heads are added on the top of $z_{inter}$ to learn the desired components (see Appendix~\ref{sec_head_arch} for
details of the head architectures).

In a symmetrical assembly, 
the ASU has at most three isologous interfaces and two heterologous interfaces (see Appendix~\ref{sec_topology}).
Each isologous interface corresponds one neighbour chain, and each heterologous interface corresponds two neighbour chains, so the ASU has at most seven neighbour chains. 
PosHead generates the relative position mas $P^{nst}$ for both isologous and heterologous interfaces:
\begin{equation}
\begin{aligned}
    P^{nst} &= \mathrm{PosHead}(z_{inter}) \\
\end{aligned}
\end{equation}

To distinguish which chain each pixel belongs to, ClusterHead predicts chain index segmentation mask:
\begin{equation}
\begin{aligned}
   H &= \mathrm{ClusterHead}(z_{inter}) \\
\end{aligned}
\end{equation}
where $H \in \mathbb{R}^{N\times N\times 7}$ indicates classification score for each neighbour chain.
$H$ also serves as confidence score map in solving symmetry axis.
As the symmetry generators differs with symmetry types, we need to predict one specific type of C, D, T, O and I:
\begin{equation}
\begin{aligned}
   Y &= \mathrm{ClsHead}(z_{inter}) \\
\end{aligned}
\end{equation}
where $Y\in \mathbb{R}^5$ is the symmetry type classification score.

\subsection{Loss functions}
\label{sec_loss}

In this section, we introduce each loss function.
The final loss is the linear combination of loss values with loss weight.

\textbf{Distogram loss.}
To supervise the learning of $z_{inter}$,
we leverage the distogram loss from AFM that utilizes a linear projection to map the inter-chain pairwise representations $z_{inter}$ onto 64 distance bins.
The bin probabilities $p^b_{ij}$ is obtained using softmax operation. 
The distogram loss is defined as the averaged cross-entropy loss:
\begin{equation}
\begin{aligned}
    L_{dist} &= -\frac{1}{N^2} \sum_{i,j}\sum_{b=1}^{64} y^b_{ij}\log p^b_{ij}
\end{aligned}
\end{equation}
where $y_{ij}$ is the ground truth bin.

\textbf{Relative position map loss.}
We assume the atoms has interaction within $d_{clamp}=20$ {\AA} distance and only consider close distance interaction.
If the distance of two $C^\alpha$ exceeds $d_{clamp}$, their loss will not be considered.
The relative position map error is penalized by an averaged L2-loss:
\begin{equation}
\begin{aligned}
    M_v &= D^{gt}<d_{clamp} \\ 
    L_{pos} &= \frac{1}{N^2}||P^{nst}*M_v - P^{nst, gt}*M_v||_F
\end{aligned}
\end{equation}

\textbf{Chain index loss.}
Considering the quotient space property, 
the desired chain index map $M$ should fulfill the following relationship:
$\forall m_i, m_j \in M$, if residues $i,j$ are from the same chain, then $m_i=m_j$.
We transfer $H$ to probability by applying softmax $H_p = \mathrm{Softmax}(H) \in \mathbb{R}^{N^2\times 7}$ and 
the chain index loss is formulated as:
\begin{equation}
\begin{aligned}
    F &= (H_p*M_v)(H_p*M_v)^T \\ 
    L_{id} &= \frac{1}{N^4}\sum_{i,j}(1-F_{ij})\cdot F^{gt}_{ij} +F_{ij}\cdot (1-F^{gt}_{ij})
\end{aligned}
\end{equation}
where the ground truth $F^{gt}_{ij}$ equals to $1$ if the chain indexes in pixel $i$ and $j$ are equal.
Since optimizing all pairs of chain indexes ($N^4$ pairs) will encounter memory issue, in practice we instead randomly sample pixels in $H$ and calculate loss on them.



\section{Experiments}

\begin{table*}[t]
\centering
\begin{tabular}{cccccc} 
\toprule
 Symmetry   &\# Train set   &\# Test set    & Avg. ASU length   & Avg. complex length \\ 
\toprule
 C      & $3,000$   & $100$     & $303$     & $778$\\
 D      & $1,000$   & $100$     & $318$     & $1,443$\\
 T      & $200$     & $20$      & $163$     & $1,955$\\
 O      & $200$     & $20$      & $200$     & $4,803$\\
 I      & $150$     & $10$      & $289$     & $17,340$\\
\bottomrule
\end{tabular}
\caption{Benchmark statistic information.}
\label{tab_data} 
\end{table*}

\begin{table*}
\centering
\scalebox{1}[1]{
\begin{tabular}{lcccccccc}
\toprule
\multirow{2}{*}{Model} & \multicolumn{2}{c}{Dimer} & \multicolumn{2}{c}{Cyclic$_{>2}$} & \multicolumn{2}{c}{Dihedral} & \multicolumn{2}{c}{Tetrahedral} \\
\cmidrule(lr){2-3}  \cmidrule(lr){4-5}  \cmidrule(lr){6-7}  \cmidrule(lr){8-9} 
& RMSD & \makecell{TM-score}  & RMSD & \makecell{TM-score} & RMSD & \makecell{TM-score} & RMSD & \makecell{TM-score} \\
\toprule
\makecell{AlphaFold-\\Multimer} 	
& 8.3   & 0.78  & 16.3  & 0.66 & 10.3 & 0.77    & 26.2 & 0.31  \\

\cdashline{2-9}
\specialrule{0em}{1pt}{1pt}
& \textcolor[RGB]{50,205,50}{0.5}   & \color{red} -0.04  & \textcolor[RGB]{50,205,50}{6.4}  & \textcolor[RGB]{50,205,50}{0.21} & \textcolor[RGB]{50,205,50}{3.2} & \textcolor[RGB]{50,205,50}{0.07}  & \textcolor[RGB]{50,205,50}{21.2} & \textcolor[RGB]{50,205,50}{0.61}  \\
\cdashline{2-9}
\specialrule{0em}{2pt}{2pt}

\makecell{SGNet}  			
& 7.8   & 0.74  & 9.9   & 0.87 & 7.1 & 0.84  & 5.1 & 0.92  \\
\bottomrule
\end{tabular}
}
\caption{Performance comparison with AlphaFold-Multimer. The performance gain is displayed in the middle of the table (Best viewed in color).}
\label{tab_com}
\end{table*}


\subsection{Experimental setup}
\textbf{Data and metrics.}
We provide a symmetrical protein complex benchmark for evaluating our proposed method. The construction of the benchmark is as follows. We retrieved experimentally resolved structural data from the PDB database for protein multimers before 2021-08-01. 
We filtered all the symmetric protein complexes, and the ground truth symmetry information is collected from the RCSB website~\footnote{\url{https://www.rcsb.org/}}. 
Owing to limitation in memory capacity, complexes with the ASU length exceeding $2,000$ are excluded in test set.
In nature, most protein complexes have C and D symmetry, while a minority possess cubic symmetry.
To balance the data distribution of different symmetry types,
we randomly select $3,500$, $1,100$, $220$, $220$, $160$ samples for C, D, T, O and I groups respectively.
The statistic information of the benchmark is summarized in Tab.~\ref{tab_data}.
The selected samples are split into train/test set by their release date,
\textit{i.e} the newer samples are in test set.
The cutoff date is 2018-04-30 which is consistent with that used in AFM pretrained model (V1 version) to ensure no information leakage from test set.
We utilize Root Mean Squared Deviation (RMSD) and Template-Modeling Score (TM-Score)~\cite{zhang2004scoring} as evaluation metrics

\begin{table*}
\centering
\begin{tabular}{cccccccc} 
\toprule
 Local Metrics   && Cyclic   & Dihedral   & Tetrahedral  & Octahedral & Icosahedral & All \\ 
\midrule
 RMSD      && 6.0    & 5.7   & 10.6  & 7.4   & 15.2  & 6.8 \\
\specialrule{0em}{1pt}{1pt} 
 TM-score  && 0.76   & 0.79  & 0.65  & 0.84  & 0.51  & 0.76 \\
\bottomrule
\end{tabular}
\caption{Results of SGNet on subsets of different symmetry types.}
\label{tab_result}
\end{table*}


\noindent\textbf{Implementation details.}
Following AlphaFold,
multiple sequence alignments and template features are searched and extracted to complement the ASU sequence. 
Before passing to the network, 
the input sequences are randomly cropped to $300$ residues in the training. 
The protein structure model is pretrained on monomer prediction task~\cite{Ahdritz2022.11.20.517210}.
The symmetry module contains 6 evoformer blocks.
It should be noted that symmetry type is optional to input,
by default no ground truth symmetry information is provided in the inference stage.
Our experiments are conducted on a server with eight Nvidia A100 GPUs, where each GPU has 40 GB memory.
The training process consists of 30 epochs.
The total batch size is 16.
During training, the parameters of the pretrained model are fixed, while the parameters of the symmetry module are learned from scratch.

\begin{figure}[t]
\centering 
\includegraphics[width=0.95\linewidth]{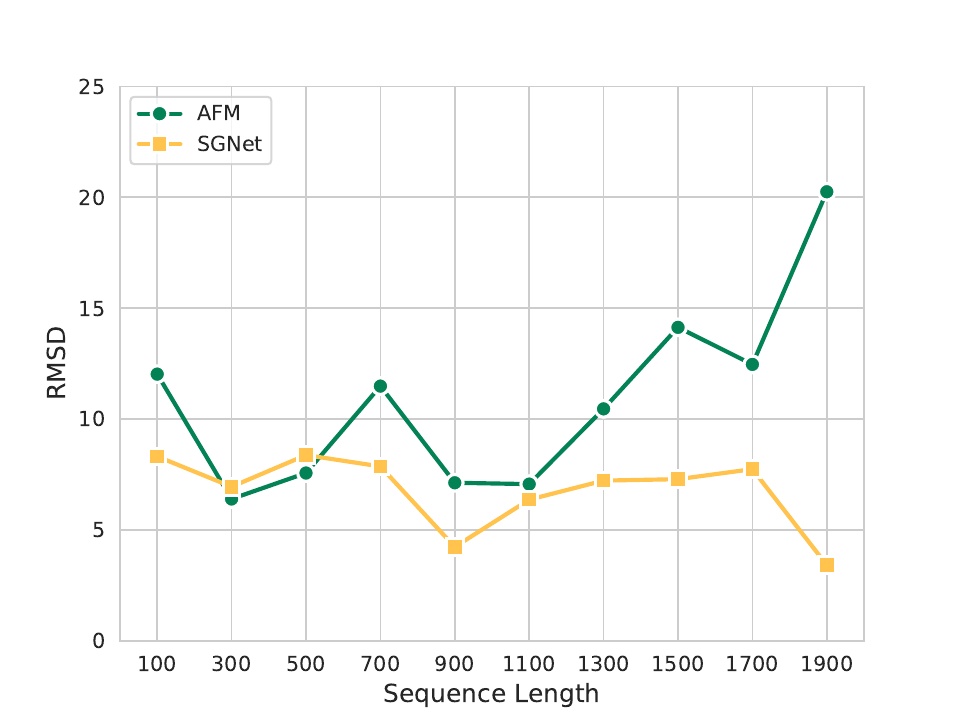}
\caption{Performance variation with respect to protein length.}
\label{fig_len}
\end{figure}

\begin{table*}[t]
\centering
\begin{tabular}{|cccc|cc|} 
 \hline
 \multicolumn{4}{|c}{Extra ground truth information} & \multicolumn{2}{|c|}{Metrics} \\
 \hline
 Symmetry type  & ASU structure   & Chain index map    & Relative position map    & RMSD   & TM-score \\ 
 \hline
 -  & - & - & -                                     & 6.8 & 0.76 \\ 
 \checkmark & - & - & -                             & 6.7 & 0.76 \\
 - & \checkmark & - & -                             & 6.7 & 0.76 \\
 \checkmark & \checkmark & \checkmark  & -          & 5.6 & 0.77 \\
 \checkmark & \checkmark & - & \checkmark           & 4.1 & 0.84 \\
 \checkmark & \checkmark & \checkmark & \checkmark  & 0.1 & 0.96 \\
 \hline
\end{tabular}
\caption{Performance gap analysis.}
\label{tab_ablation}
\end{table*}

\begin{figure*}[t]
\centering 
\includegraphics[width=0.9\textwidth]{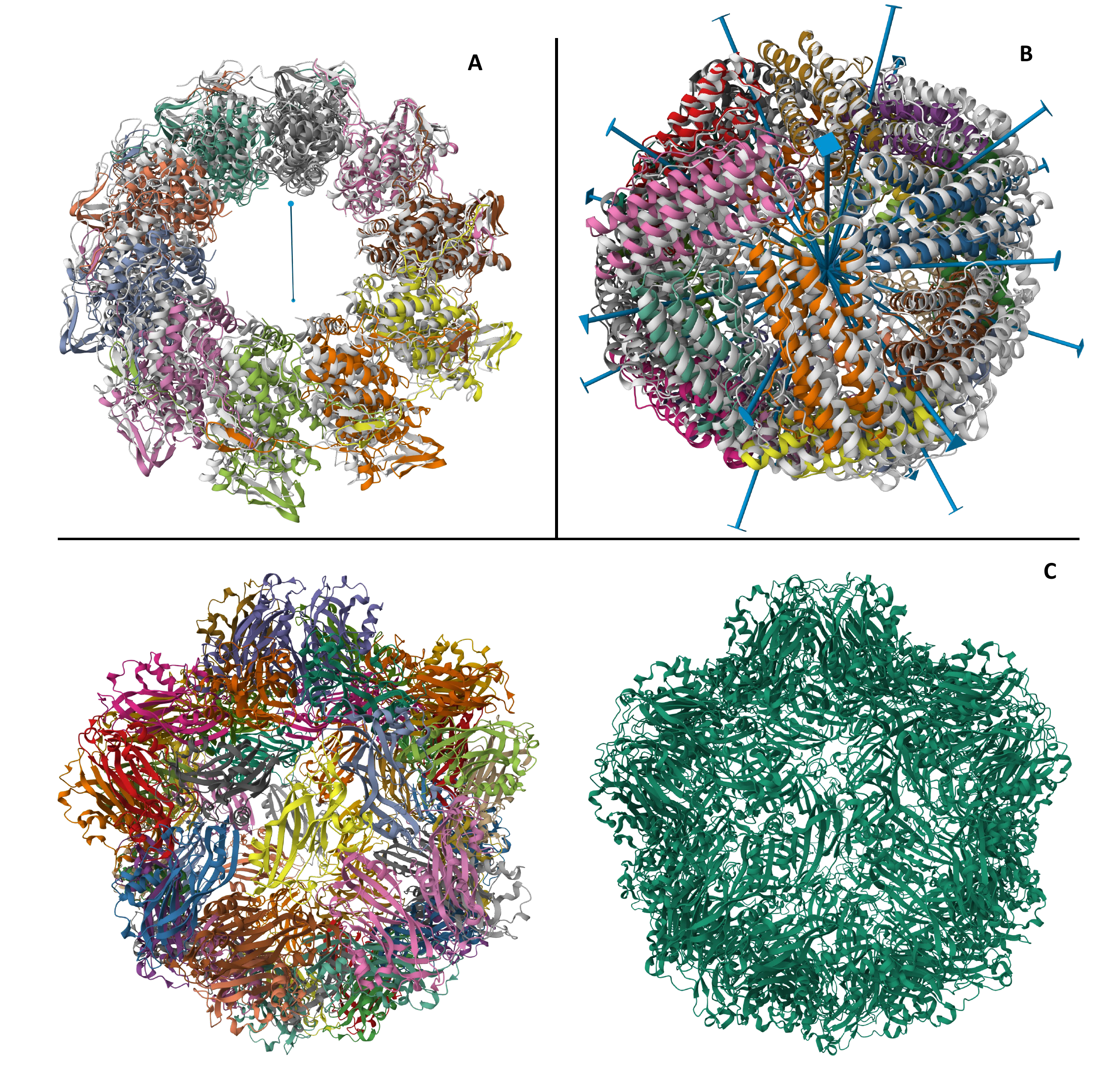}
\caption{(A) and (B) are protein superimposition where the blue symmetry axes are also rendered. In (C) we place the ground truth on right for better visualization. (A) 2qvj, C10 symmetry. (B) 5xx9, O symmetry. (C) 5yl1, I symmetry.}
\label{fig_example}
\end{figure*}

\subsection{Comparison results}
Since AlphaFold-Multimer has cubic complexity with sequence length,
if the input sequence is very long, it fails on single GPU evaluation due to memory limitation.
On the other hand,  the computational load of our method mainly depends on the length of the ASU, with SGNet demonstrating the capability to infer across all test set samples. 
For a fair comparison with AFM,
in Tab.~\ref{tab_com} we report the performance of samples which are successfully dissolved by AFM on single GPU. 
The count of successfully resolved samples in the C, D, and T symmetry categories are 72, 69, and 4, respectively.
In the table, symmetric dimers extracted from the C symmetry group are itemized separately for a detailed comparison,
and the notation $>2$ indicates the samples comprise more than two monomers. 
The data presented in the table elucidates that 
our method outperforms the baseline across all symmetry subsets, while the performance on dimers is comparable.
Both methods exhibit proficiency in generating accurate symmetric assemblies with sequences of relatively modest length. 
However, AFM's prediction accuracy for cubic symmetry is markedly deficient. 
In contrast, SGNet shows promise in predicting cubic symmetry within longer sequences, particularly for exceedingly large protein complexes.

\subsection{Influence of protein sequence length}
Protein sequence length is one crucial factor that influences the structure prediction accuracy of AFM.
This assertion could be further corroborated within the realm of symmetric protein prediction. 
In the Fig.~\ref{fig_len}, we depict the RMSD value variance with respect to sequence length.
To make the plot clearer, we divide the data into several bins based on sequence length and calculate the average RMSD for each group. 
The figure shows that SGNet is more robust to the protein length change, thanks to our two-stage generation process.

\subsection{Full evaluation of SGNet}
As mentioned before, proteins with cubic symmetries are usually very large. 
Render their full-scale evaluation is challenging, if not unfeasible, particularly when we consider all atoms and all chain permutations. 
To ease the metric computation, we restrict our evaluation to a select subset of monomers that are essential for constituting the complete assembly (e.g. for icosahedra, the subunits at the five-fold, three-fold, and two-fold symmetry axes) to reduce the computational cost~\cite{watson2023novo}.
Concretely, the local subset comprises 2, 3, 4, 4, 5, and 6 monomers for symmetries C$2$, C$n$, D$n$, T, O, and I, respectively.
It should be noted the local metrics and global metrics are consistent in assessing the quality of predicted proteins. 
This strict correlation comes from the perfect symmetry in our output protein complexes.
The full evaluation results on the test set of our method are presented in Tab.~\ref{tab_result}, where the metrics are calculated locally.
We show some cases in Fig.~\ref{fig_example}.
From the results we could conclude that our method performance well on C and D symmetry. 
However, the construction of cubic symmetry remains a formidable challenge, primarily due to the necessity of inferring multiple interfaces, which makes it more sensitive to possible errors in head prediction.

\subsection{Ablation study}
To investigate the performance gap and analyze the effectiveness of the components in symmetry module, we replace some of the predicted outputs with the actual ground truth values and then look at how this affect the structures.
The evaluation results are shown in Tab.~\ref{tab_ablation}.
The first three rows of the table show that correctly predicting the ASU structure and the type of symmetry have little effect on the overall performance, suggesting these parts of our method are learned well.
In fact, the chain index head and relative position map are more important since they are directly related to geometric information, and they are harder to predict accurately.
The last row of the table, where we replaced everything with ground truth values, shows the systematic error of our framework in the implementation.

\section{Conclusion}
Most large proteins exhibit symmetry in their structure.
In this paper, our aim is to investigate protein–protein interactions within symmetry assemblies and present an effective framework, SGNet, for predicting symmetrical protein folding.
Our main focus is on modeling symmetrical structures and we accomplish this by proposing the use of a relative position map to streamline the representation of symmetry relations.
Through careful modeling of symmetry, our framework can capture all global symmetry types in predicting quaternary protein structures.
Additionally, SGNet conducts feature extraction on a single subunit and leverages our proposed symmetry module to generate the entire assembly. This approach greatly reduces the computational complexity caused by lengthy sequences.
Our future work will focus on improving performance of structure prediction for cubic symmetry.
{
    \small
    \bibliographystyle{ieeenat_fullname}
    \bibliography{main}
}

\clearpage
\appendix

\section*{Appendix}

\begin{table*}[ht]
\caption{Structural topologies.}
\label{tab_topology}
\centering
\begin{tabular}{|c|c|c|c|} 
 \hline
 Symmetry Type  & Isologous   & Heterologous    & remark on $n$ \\ 
 \hline
 C         & 1     & 0         & $n=2$\\
 C         & 0     & 1         & $n>2$\\
 D         & 3     & 0         & $n=4$\\
 D         & 1     & 1         & $n>4$\\
 D         & 2     & 1         & $n>4$\\
 T         & 0     & 2         & $n\geq 12$\\
 T         & 1     & 2         & $n\geq 12$\\
 T         & 1     & 1         & $n\geq 12$\\
 O         & 0     & 2         & $n\geq 24$\\
 O         & 1     & 2         & $n\geq 24$\\
 O         & 1     & 1         & $n\geq 24$\\
 I         & 0     & 2         & $n\geq 60$\\
 I         & 1     & 2         & $n\geq 60$\\
 I         & 1     & 1         & $n\geq 60$\\
 \hline
\end{tabular}
\end{table*}

\begin{figure*}[h]
\centering
\includegraphics[width=0.8\textwidth]{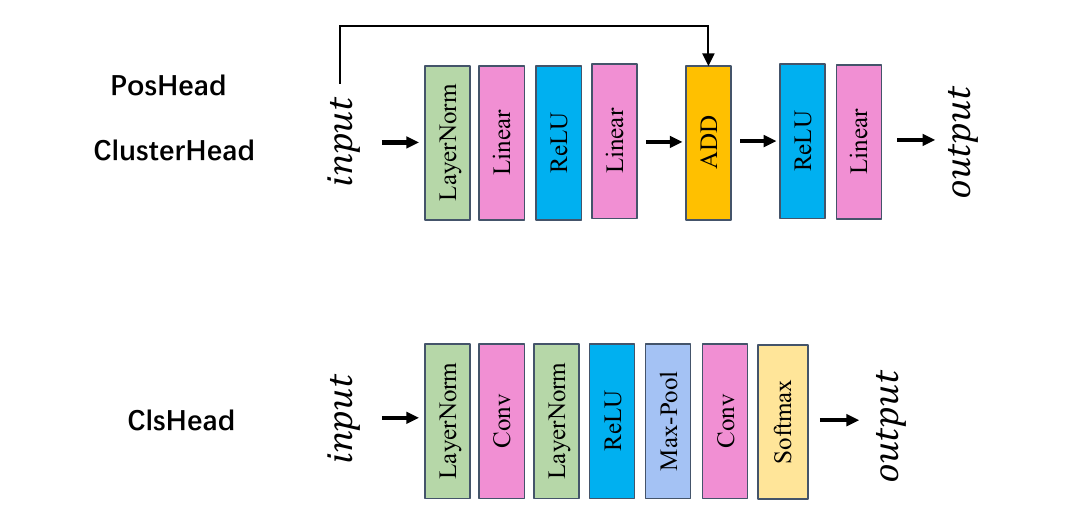}
\caption{Neural architecture of heads.}
\label{fig_arch}
\end{figure*}

\section{Asymmetric units}
The asymmetric unit (ASU) is the smallest portion of a protein assembly structure to which symmetry operations can be applied in order to generate the complete structure. 
ASU is composed of several protein chains in the assembly and can be determined directly from the input sequence~\cite{johansson2017automatic}. 
For instance, the subunit of a sample with sequence $\{A_1A_2A_3A_4B_1B_2\}$ is $\{A_1A_2B_1\}$.

\section{Structural topology}
\label{sec_topology}

We list all cases of structural topologies in Tab.~\ref{tab_topology}.
It should be noted that real protein complexes can have more interfaces~\cite{ahnert2015principles} compared to our modeling.
However, it is possible to establish a direct correlation between actual protein structures and the aforementioned idealized modelling by regarding certain weaker inter-subunit contacts as circumstantial.

For the two subunits of one interface, their relative distance maps $D_{1,2}, D_{2,1} \in \mathbb{R}^{N\times N}$ satisfy following properties:
\begin{itemize}
    \item Isologous: $D_{1,2}=D_{2,1}$, $D_{1,2}$ is a symmetric matrix
    \item Heterologous: $D_{1,2}^T=D_{2,1}$, $D_{1,2}$ is not symmetric
\end{itemize}

\section{Quotient space property} 
\label{sec_quotient}
The symmetric property poses an inevitable issue to the label assignment since it is not unique. 
Specifically, swapping the index between any two target positions will still lead to a valid assignment and a identical multimer structure. That is to say, the label assignment is permutation invariant which is referred to the quotient space property.
Instead of using implicit positional encoding~\cite{Richard2021alphafoldmultimer},
we explicitly address the problem by modeling symmetry.




\section{Implemenation details}
\subsection{Network architecture}
\label{sec_head_arch}
PosHead and ClusterHead have same architecture which consists of one residual structure.
The channels are the same as input tensor $z_{inter}$ which is 128 for our implementation.
ClsHead is a CNN that predicts a classification score over 5 possible symmetry types.
The channels of two covolution layers are 128 and 64.
The architectures are plotted in Fig.~\ref{fig_arch}.

\subsection{Training and inference}
The channels of single representation and pair representation are 384 and 128 respectively. The number of selection is 30 in the symmetry generator.
The total loss is the linear combination of loss values with loss weights:
\begin{equation}
\begin{aligned}
    L = \lambda_1 L_{pos} + \lambda_2 L_{dist} + \lambda_3 L_{id} + \lambda_4 L_{cls}
\end{aligned}
\end{equation}
We use Adam optimizer with $\beta_1=0.9, \beta_2=0.999$ to optimize the network.
Regarding the loss weights, $\lambda_{1\sim 4}$ are $0.5, 0.3, 0.3, 0.1$ respectively. 
The learning rate warms up from 0 to 1e-3 for the first 1,000 steps. 
After that, the learning rate decays by 5\% every 5,000 steps. 

\section{Symmetry generator}
\label{sec_sym_generator}

In this section, 
we provide Python pseudo code for all symmetry generators.
We force the output structure to be strictly symmetrical and consider different cases.
For example, one heterologous interface plus isologous interface are sufficient to reconstruct the whole assembly for $D$ symmetry .  
If a protein complex has one heterologous interface and two isologous interfaces, then we choose the isologous interface which has more contact points to calculate the symmetry axis.
For reconstructing neighbour chains by contact interfaces, we adopt Kabsch algorithm~\cite{kabsch1976solution}.

\begin{algorithm}[h]
\caption{Construct local frame from 3 points}
\label{alg_3points}
    \begin{algorithmic}[1]
    
    \renewcommand{\algorithmicrequire}{\textbf{Input:}}
    \renewcommand{\algorithmicensure}{\textbf{Output:}}
    \Require Three points $x_1, x_2, x_3 \in \mathbb{R}^3$. 
    \Ensure Local frame $T \in \mathrm{SE(3)}$. 

    \renewcommand{\algorithmicrequire}{\textcolor[RGB]{50,140,10}{def}} 
    \Require Frame3Points($x_1$, $x_2$, $x_3$) 
    
    \State $v_1 = x_3-x_2$ 
    \State $v_2 = x_1-x_2$ 
    \State $e_1 = \frac{e_1}{||e_1||}$ 
    \State $u_2 = v_2 - e_1(e_1^Tv_2)$ 
    \State $e_2 = \frac{e_2}{||e_2||}$ 
    \State $e_3 = e_1\times e_2$ 
    \State $R = \mathrm{concat}(e_1,e_2,e_3)$  \Comment{$R\in \mathbb{R}^{3\times 3}$}
    \State $t = x_2$
    \State $T = (R,t)$ 
    \State \Return $T$
    \end{algorithmic}
\end{algorithm}

\begin{algorithm}[h]
\caption{Calculate the vector from a point to a line}
\label{alg_vec2line}
    \begin{algorithmic}[1]
    
    \renewcommand{\algorithmicrequire}{\textbf{Input:}}
    \renewcommand{\algorithmicensure}{\textbf{Output:}}
    \Require $p \in \mathbb{R}^3$, $l \in \mathbb{R}^{2\times 3}$. 
    \Ensure $v \in \mathbb{R}^3$. 

    \renewcommand{\algorithmicrequire}{\textcolor[RGB]{50,140,10}{def}} 
    \Require VecPoint2Line($p$, $l$) 
    
    \State $s, e = l$
    \State $s = s - p$
    \State $e = e - p$
    \State $v_1 = e - s$
    \State $k = -\frac{s^Tv_1}{||v_1||^2}$
    \State $v =s + kv_1$
    \State \Return $v$
    \end{algorithmic}
\end{algorithm}

\begin{algorithm}[h]
\caption{Apply rotational symmetry operation}
\label{alg_apply_sym}
    \begin{algorithmic}[1]
    \renewcommand{\algorithmicrequire}{\textbf{Input:}}
    \renewcommand{\algorithmicensure}{\textbf{Output:}}
    \Require Point set $p$, Symmetry axis $l$, Symmetry order $n$. 
    \Statex
    \Comment{$p \in \mathbb{R}^{N\times 3}$, $l \in \mathbb{R}^{2\times 3}$, $n \in \mathbb{N}$} 
    \Ensure All atom positions $p_{all}$.
    \Statex
    \Comment{$p_{all}\in \mathbb{R}^{n\times N \times 3}$}
    
    \renewcommand{\algorithmicrequire}{\textcolor[RGB]{50,140,10}{def}}
    \Require ApplySymmetryOp($p$, $l$, $n$)

    \State $v = \mathrm{VecPoint2Line}([0,0,0], l)$
    \State $p = p - v$
    \State $l_n = \frac{l[0]-l[1]}{||l[0]-l[1]||}$
    \State $l_x, l_y, l_z = l_n$
    \State $C = \left[ \begin{array}{ccc}
                        0 & -l_z & l_y \\
                        l_z & 0 & -l_x \\
                        -l_y & l_x & 0
                        \end{array} \right ]$
                        
    \For{$i \in \{0,1,...,n-1\}$}
    \State $\theta = \frac{2\pi i}{n}$
    \State $R = \left[ \begin{array}{ccc}
                        1 & 0 & 0 \\
                        0 & 1 & 0 \\
                        0 & 0 & 1
                        \end{array} \right ]$ 
            + $(\sin\theta) C$
            + $(1-\cos\theta)C^TC$
    \State $p_i = Rp$
    \State $p_i = p_i + v$
    \EndFor

    
    \State \Return $p_{all}$
    \end{algorithmic}
\end{algorithm}

\begin{algorithm}[h]
\caption{Reconstruct neighbour chain}
\label{alg_re_nchain}
    \begin{algorithmic}[1]
    
    \renewcommand{\algorithmicrequire}{\textbf{Input:}}
    \renewcommand{\algorithmicensure}{\textbf{Output:}}
    \Require Backbone frames $\mathcal{T}$, Relative position map $P^r$, Point index set $S$.
    \Statex
    \Comment{$\mathcal{T} \in \mathbb{R}^{N\times 4\times 4}$, $P^r \in \mathbb{R}^{N\times N\times 3}$, $S \in \mathbb{R}^{K\times 2}$} 
    \Ensure $C^{\alpha}$ coordinate of the neighbour chain $p$. 
    \Statex
    \Comment{$p \in \mathbb{R}^{N\times 3}$}

    \renewcommand{\algorithmicrequire}{\textcolor[RGB]{50,140,10}{def}}
    \Require ReconstructNeighbourChain($\mathcal{T}$, $P^r$, $S$) 

    \State $K = \mathrm{len}(S)$
    \For {$i \in \{0,1,...,K-1\}$}
    \State $s, e = S[i]$ 
    \State $T_s = \mathcal{T}[s]$ 
    \State $T_e = \mathcal{T}[e]$ 
    \State $x_i = T_s\circ P^r[s, e]$   \Comment{$x_i \in \mathbb{R}^3$}
    \State $y_i = T_e.t$                \Comment{$y_i \in \mathbb{R}^3$}
    \EndFor
    
    \State Solve $T^*=\mathop{\min}\limits_{T} \sum_i ||y_i-T\circ x_i||$\ by Kabsch algorithm
    \For {$i \in \{0,1,2,...,N-1\}$}
    \State $p[i] = T^*\circ {T_i.t}$  
    \EndFor
    
    \State \Return $p$
    \end{algorithmic}
\end{algorithm}

\begin{algorithm}[h]
\caption{Calculate symmetry axis of isologous interface}
\label{alg_iso2axis}
    \begin{algorithmic}[1]
    \renewcommand{\algorithmicrequire}{\textbf{Input:}}
    \renewcommand{\algorithmicensure}{\textbf{Output:}}
    \Require Backbone frames $\mathcal{T}$, Relative position map $P^r$, Point index set $S$. 
    \Statex
    \Comment{$\mathcal{T} \in \mathbb{R}^{N\times 4\times 4}$, $P^r \in \mathbb{R}^{N\times N\times 3}$, $S \in \mathbb{R}^{K\times 2}$} 
    \Ensure Symmetry axis $l$.
    \Statex
    \Comment{$l \in \mathbb{R}^{2\times 3}$}

    \renewcommand{\algorithmicrequire}{\textcolor[RGB]{50,140,10}{def}}
    \Require SymmetryAxisIsologous($\mathcal{T}$, $P^r$, $S$) 

    \State $K = \mathrm{len}(S)$
    \For {$i \in \{0,1,...,K-1\}$}
    \State $s, e = S[i]$ 
    \State $T_s = \mathcal{T}[s]$ 
    \State $T_e = \mathcal{T}[e]$ 
    \State $x_i = T_s\circ P^r[s, e]$  \Comment{$x_i \in \mathbb{R}^3$}
    \State $y_i = T_e.t$               \Comment{$y_i \in \mathbb{R}^3$}
    \State $m_i = \frac{x_i+y_i}{2}$ 
    \EndFor
    \State Solve $l=\mathop{\min}\limits_{l^{\prime}} \sum_i dist(l^{\prime},m_i)$ by least square algorithm. 
    \State \Return $l$
    \end{algorithmic}
\end{algorithm}

\begin{algorithm}[h]
\caption{Calculate symmetry axis of heterologous interface}
\label{alg_hetero2axis}
    \begin{algorithmic}[1]
    \renewcommand{\algorithmicrequire}{\textbf{Input:}}
    \renewcommand{\algorithmicensure}{\textbf{Output:}}
    \Require $C^{\alpha}$ coordinate of base ASU $p_1$, $C^{\alpha}$ coordinate of neighbour chains $p_2$, $p_3$, Number of replicates $n$.
    \Statex
    \Comment{$p_1, p_2, p_3 \in \mathbb{R}^{N\times 3}$} 
    \Ensure Symmetry axis $l$.
    \Statex
    \Comment{$l \in \mathbb{R}^{2\times 3}$}
    
    \renewcommand{\algorithmicrequire}{\textcolor[RGB]{50,140,10}{def}}
    \Require SymmetryAxisHeterologous($p_1$, $p_2$, $p_3$) 
    
    \State $c_1 = \mathrm{mean}(p_1)$ \Comment{$c_1 \in \mathbb{R}^3$}
    \State $c_2 = \mathrm{mean}(p_2)$ \Comment{$c_2 \in \mathbb{R}^3$}
    \State $c_3 = \mathrm{mean}(p_3)$ \Comment{$c_3 \in \mathbb{R}^3$}

    \If{$n==3$}
    \State $\theta_{rot} = \pi - \frac{2\pi}{n}$ 
    \Else
    \State $\theta_{rot} = \frac{2\pi}{n}$ 
    \EndIf
    
    \State $T_{rot} = \mathrm{Frame3Points}(c2, c1, c3)$ 
    \State $c_{2,local}=T_{rot}^{-1}\circ c_2$
    \State $c_{3,local}=T_{rot}^{-1}\circ c_3$
    \State $d_p = c_{3,local}-c_{2,local}$  
    \State $r = \frac{||d_p||}{2\sin(\theta_{rot})}$ 

    \State $d_{o1} = [-d_p[1], d_p[0]]$  \Comment{$d_{o1}^Td_p=0$}
    \State $d_{o2} = [d_p[1], -d_p[0]]$  \Comment{$d_{o2}^Td_p=0$}
    \If{$c_{3,local}^Td_{o1}>0$}
    \State $d_{o} = \frac{d_{o1}}{||d_{o1}||}$ 
    \Else
    \State $d_{o} = \frac{d_{o2}}{||d_{o2}||}$ 
    \EndIf

    \State $o_{local} = r d_o$ 
    \State $z_{local} = [0,0,1]$ 

    \State $o = T_{rot} \circ o_{local}$ 
    \State $z = T_{rot} \circ z_{local}$ 

    \State $l = [o, z]$ 
    
    \State \Return $l$
    \end{algorithmic}
\end{algorithm}

\begin{algorithm}[h]
\caption{Select anchor points for reconstructing neighbour chain}
\label{alg_pos_idx}
    \begin{algorithmic}[1]
    \renewcommand{\algorithmicrequire}{\textbf{Input:}}
    \renewcommand{\algorithmicensure}{\textbf{Output:}}
    \Require Chain id map $M$, confidence score map $C$, Number of selection $K$.
    \Statex
    \Comment{$M \in \mathbb{R}^{7\times N\times N}$, $C \in \mathbb{R}^{7\times N\times N}$} 
    \Ensure Point index set $S$.
    \Statex
    \Comment{$S \in \mathbb{R}^{7\times K\times 2}$}
    
    \renewcommand{\algorithmicrequire}{\textcolor[RGB]{50,140,10}{def}}
    \Require GetPointIndexSet($M$, $C$, $K$) 

    \State $M_{value} = \left[1, 2, 3, 1, 2, 1, 2\right]$

    \For {$i \in \{0,1,2,3,4,5,6\}$}
    \State $m_{valid} = M[i]==M_{value}[i]$
    \State $C[i][\sim m_{valid}] = +\infty$
    \State $idx = \mathop{\arg\max}_i C[i]$
    \State $S[i] = idx[:K]$
    \EndFor

    \State \Return $S$
    \end{algorithmic}
\end{algorithm}


\begin{algorithm}[h]
\caption{Cyclic symmetry generator}
\label{alg_c_sym}
    \begin{algorithmic}[1]
    \renewcommand{\algorithmicrequire}{\textbf{Input:}}
    \renewcommand{\algorithmicensure}{\textbf{Output:}}
    \Require Backbone frames $\mathcal{T}$, Confidence score map $C$, Nearest position map $P^{nst}$, Chain id map $M$, Number of selection $K$, Number of replicates $n$.
    \Statex
    \Comment{$P^{nst} \in \mathbb{R}^{N\times N}$, $M \in \mathbb{R}^{7\times N\times N}$, $C \in \mathbb{R}^{7\times N\times N}$} 
    \Ensure Atom positions of backbone $p_{all}$.
    \Statex
    \Comment{$p_{all} \in \mathbb{R}^{n\times N\times3 \times 3}$}

    \renewcommand{\algorithmicrequire}{\textcolor[RGB]{50,140,10}{def}}
    \Require CyclicSymmetryGenerator($\mathcal{T}$, $P^{nst}$, $M$, $C$, $K$, $n$) 

    \State $p_{bb} := \text{backbone atom positions}$ \Comment{$p_{bb} \in \mathbb{R}^{N\times 3\times3 }$}    
    \State $p_0 := C^{\alpha} \text{ position}$ \Comment{$p_0 \in \mathbb{R}^{N\times 3 }$}
    \State $S = \mathrm{GetPointIndexSet}(M, C, K)$ \Comment{$S \in \mathbb{R}^{7\times K\times 2}$}
    
    \If {$n==2$} 
    \State $l = \mathrm{SymmetryAxisIsologous}(\mathcal{T}, P^{nst}, S[0])$ 
    \State $p_{all} = \mathrm{ApplySymmetryOp}(p_0, l, 2)$ 
    
    \Else 
    \State $p_1 = \mathrm{ReconstructNeighbourChain}(\mathcal{T}, P^{nst}, S[3])$
    \State $p_2 = \mathrm{ReconstructNeighbourChain}(\mathcal{T}, P^{nst}, S[4])$
    \State $l = \mathrm{SymmetryAxisHeterologous}(p_0, p_1, p_2)$ 
    \State $p_{all} = \mathrm{ApplySymmetryOp}(p_{bb}, l, n)$ 
    \EndIf
    \State \Return $p_{all}$
    \end{algorithmic}
\end{algorithm}



\begin{algorithm*}[h]
\caption{Make predicted symmetry axes consistent in D2 symmetry}
\label{alg_isosceles2axis}
    \begin{algorithmic}[1]
    \renewcommand{\algorithmicrequire}{\textbf{Input:}}
    \renewcommand{\algorithmicensure}{\textbf{Output:}}
    \Require $C^{\alpha}$ coordinate of base ASU $p_0$, $C^{\alpha}$ coordinate of neighbour chains $p_1$, $p_2$, $p_3$.
    \Statex
    \Comment{$p_1, p_2, p_3 \in \mathbb{R}^{N\times 3}$} 
    \Ensure Symmetry axes $l_1$, $l_2$, $l_3$.
    \Statex
    \Comment{$l_1, l_2, l_3 \in \mathbb{R}^{2\times 3}$}

    \renewcommand{\algorithmicrequire}{\textcolor[RGB]{50,140,10}{def}}
    \Require SymmetryAxisIsosceles($p_0$, $p_1$, $p_2$, $p_3$)

    \State $O = \mathrm{mean}(p_0)$
    \State $A = \mathrm{mean}(p_1)$
    \State $B = \mathrm{mean}(p_2)$
    \State $C = \mathrm{mean}(p_3)$

    \State $x_{neg} = O+O-B$
    \State $T_{g} = \mathrm{Frame3Points}(x_{neg}, O, A)$

    \State $OA = \frac{||O-A||+||B-C||}{2}$
    \State $OB = \frac{||O-B||+||A-C||}{2}$
    \State $OC = \frac{||O-C||+||A-B||}{2}$

    \State $a_2 = OA^2$
    \State $b_2 = OB^2$
    \State $c_2 = OC^2$    
    \Statex
    \Comment{Solve tetrahedron volume and triangle area by Heron's formula}
    \State $V = \sqrt{(a_2+b_2-c_2)(a_2-b_2+c_2)(-a_2+b_2+c_2)/72}$
    \State $S = \sqrt{(OA+OB+OC)(OA+OB-OC)(OA-OB+OC)(-OA+OB+OC)}/4$

    \State $\theta_{OB} = \arcsin(\frac{3\cdot V\cdot OB}{2S^2})$

    \State $\theta_{AOB} = \arcsin((b_2+a_2-c_2)/(2\cdot OB\cdot OA))$ 
    \State $\theta_{ABO} = \arcsin((b_2+c_2-a_2)/(2\cdot OB\cdot OC))$

    \State $A_{local\_x} = OA\cdot\cos(\theta_{AOB})$
    \State $A_{local\_y} = OA\cdot\sin(\theta_{AOB})$
    \State $B_{local\_x} = OB$
    \State $C_{local\_x} = OC\cdot\cos(\theta_{ABO})$
    \State $C_{local\_y} = OC\cdot\sin(\theta_{ABO})$

    \State $A_{local} = [A_{local\_x}, A_{local\_y}, 0]$
    \State $B_{local} = [B_{local\_x}, 0, 0]$
    \State $C_{local} = [C_{local\_x}, C_{local\_y}, 0]$
    \State $O_{local} = [0, 0, 0]$

    \State $R = \left[ \begin{array}{ccc}
                        1 & 0 & 0 \\
                        0 & \cos(\theta_{OB}) & -\sin(\theta_{OB}) \\
                        0 & \sin(\theta_{OB}) & \cos(\theta_{OB})
                        \end{array} \right ]$ 
    \State $T_{rot} = (R, 0)$

    \State $C_{local}$ = $T_{rot}\circ C_{local}$

    \State $O_{g} = T_{g}\circ O_{local}$
    \State $A_{g} = T_{g}\circ A_{local}$
    \State $B_{g} = T_{g}\circ B_{local}$
    \State $C_{g} = T_{g}\circ C_{local}$

    \State $OA_2 = \frac{O_g + A_g}{2}$
    \State $OB_2 = \frac{O_g + B_g}{2}$
    \State $OC_2 = \frac{O_g + C_g}{2}$
    \State $AB_2 = \frac{A_g + B_g}{2}$
    \State $AC_2 = \frac{A_g + C_g}{2}$
    \State $BC_2 = \frac{B_g + C_g}{2}$
    
    \State $l_1 = [OA_2, BC_2]$
    \State $l_2 = [OB_2, AC_2]$
    \State $l_3 = [OC_2, AB_2]$

    \State \Return $l_1$, $l_2$, $l_3$
    \end{algorithmic}
\end{algorithm*}

\begin{algorithm}[h]
\caption{Make predicted symmetry axes consistent in Dn symmetry}
\label{alg_dn2axis}
    \begin{algorithmic}[1]
    \renewcommand{\algorithmicrequire}{\textbf{Input:}}
    \renewcommand{\algorithmicensure}{\textbf{Output:}}
    \Require $C^{\alpha}$ coordinate of base ASU $p_0$, Symmetry axes $l_1$, $l_2$, Number of replicates $n$.
    \Statex
    \Comment{$l_1, l_2 \in \mathbb{R}^{2\times 3}$}
    \Ensure Symmetry axes $l_3$, $l_4$.
    \Statex
    \Comment{$l_3, l_4 \in \mathbb{R}^{2\times 3}$}

    \renewcommand{\algorithmicrequire}{\textcolor[RGB]{50,140,10}{def}}
    \Require SymmetryAxisIso2($p_0$, $l_1$, $l_2$, $n$)

    \State $l_3 = \mathrm{ApplySymmetryOp}(l_1, l_2, 2)$ 
    \State $l_4 = \mathrm{ApplySymmetryOp}(l_2, l_1, 2)$ 
    
    \State $p_1 = \mathrm{ApplySymmetryOp}(p_0, l_1, 2)$ 
    \State $p_2 = \mathrm{ApplySymmetryOp}(p_0, l_2, 2)$ 
    \State $p_3 = \mathrm{ApplySymmetryOp}(p_1, l_4, 2)$ 
    \State $p_4 = \mathrm{ApplySymmetryOp}(p_2, l_3, 2)$ 
    
    \State $l_5 = \mathrm{SymmetryAxisHeterologous}(p_0, p_3, p_4, n/2)$
    \State $c_0 = \mathrm{mean}(p_0)$
    \State $c_1 = \mathrm{mean}(p_1)$
    \State $c_2 = \mathrm{mean}(p_2)$

    \If{$||c_0-c_1||<||c_0-c_2||$}
    \State $m = \frac{c_0+c_1}{2}$
    \Else
    \State $m = \frac{c_0+c_2}{2}$
    \EndIf

    \State $e = \mathrm{VecPoint2Line}(m, l_5)$
    \State $l_6 = [m, m+e]$
    
    \State \Return $l_5$, $l_6$
    \end{algorithmic}
\end{algorithm}

\begin{algorithm}[h]
\caption{Dihedral symmetry generator}
\label{alg_d_sym2}
    \begin{algorithmic}[1]
    \renewcommand{\algorithmicrequire}{\textbf{Input:}}
    \renewcommand{\algorithmicensure}{\textbf{Output:}}
    \Require Backbone frames $\mathcal{T}$, Confidence score map $C$, Nearest position map $P^{nst}$, Chain id map $M$, Number of selection $K$, Number of replicates $n$.
    \Statex
    \Comment{$P^{nst} \in \mathbb{R}^{N\times N}$, $M \in \mathbb{R}^{7\times N\times N}$, $C \in \mathbb{R}^{7\times N\times N}$} 
    \Ensure Atom positions of backbone $p_{all}$.
    \Statex
    \Comment{$p_{all} \in \mathbb{R}^{n\times N\times 3\times 3}$}

    \renewcommand{\algorithmicrequire}{\textcolor[RGB]{50,140,10}{def}}
    \Require DihedralSymmetryGenerator($\mathcal{T}$, $P^{nst}$, $M$, $C$, $K$, $n$) 
    \State $p_{bb} := \text{backbone atom positions}$ \Comment{$p_{bb} \in \mathbb{R}^{N\times 3\times3 }$}    
    \State $p_0 := C^{\alpha} \text{ position}$ \Comment{$p_0 \in \mathbb{R}^{N\times 3 }$}
    \State $S = \mathrm{GetPointIndexSet}(M, C, K)$ \Comment{$S \in \mathbb{R}^{7\times K\times 2}$}
    \State $iso_2 = \sum S[1]$
    \State $hetero_2 = \sum S[3]$
    
    \If {$n==4$} 
    \State $l_1 = \mathrm{SymmetryAxisIsologous}(\mathcal{T}, P^{nst}, S[0])$ 
    \State $l_2 = \mathrm{SymmetryAxisIsologous}(\mathcal{T}, P^{nst}, S[1])$ 
    \State $l_3 = \mathrm{SymmetryAxisIsologous}(\mathcal{T}, P^{nst}, S[2])$ 
    \State $p_{1} = \mathrm{ApplySymmetryOp}(p_0, l_1, 2)$ 
    \State $p_{2} = \mathrm{ApplySymmetryOp}(p_0, l_2, 2)$ 
    \State $p_{3} = \mathrm{ApplySymmetryOp}(p_0, l_3, 2)$ 
    \State $l_1, l_2, l_3 = \mathrm{SymmetryAxisIsosceles}(p_0, p_1, p_2, p_3)$ 
    \State $p_{1} = \mathrm{ApplySymmetryOp}(p_{bb}, l_1, 2)$ 
    \State $p_{2} = \mathrm{ApplySymmetryOp}(p_{bb}, l_2, 2)$ 
    \State $p_{3} = \mathrm{ApplySymmetryOp}(p_{bb}, l_3, 2)$ 
    \State $p_{all} = \mathrm{concat}(p_{bb}, p_1, p_2, p_3)$
    
    \ElsIf {$iso_2>hetero_2$} 
    \State $l_1 = \mathrm{SymmetryAxisIsologous}(\mathcal{T}, P^{nst}, S[0])$ 
    \State $l_2 = \mathrm{SymmetryAxisIsologous}(\mathcal{T}, P^{nst}, S[1])$ 
    \State $p_{1} = \mathrm{ApplySymmetryOp}(p_0, l_1, 2)$ 
    \State $p_{2} = \mathrm{ApplySymmetryOp}(p_0, l_2, 2)$ 
    \State $l_1, l_2 = \mathrm{SymmetryAxisIso2}(p_0, l_1, l_2, n)$ 
    \State $p_2 = \mathrm{ApplySymmetryOp}(p_{bb}, l_1, n/2)$
    \State $p_{all} = \mathrm{ApplySymmetryOp}(p_2, l_2, 2)$
    
    \Else 
    \State $p_1 = \mathrm{ReconstructNeighbourChain}(\mathcal{T}, P^{nst}, S[3])$
    \State $p_2 = \mathrm{ReconstructNeighbourChain}(\mathcal{T}, P^{nst}, S[4])$
    \State $l_1 = \mathrm{SymmetryAxisHeterologous}(p_0, p_1, p_2)$ 
    \State $l_2 = \mathrm{SymmetryAxisIsologous}(\mathcal{T}, P^{nst}, S[0])$ 
    \State $p_2 = \mathrm{ApplySymmetryOp}(p_{bb}, l_1, n/2)$
    \State $p_{all} = \mathrm{ApplySymmetryOp}(p_2, l_2, 2)$
    \EndIf
    \State \Return $p_{all}$
    \end{algorithmic}
\end{algorithm}

\begin{algorithm*}[h]
\caption{Generate the whole assembly by interfaces for tetrahedral symmetry}
\label{alg_apply_tetrahedral_symmetry}
    \begin{algorithmic}[1]
    \renewcommand{\algorithmicrequire}{\textbf{Input:}}
    \renewcommand{\algorithmicensure}{\textbf{Output:}}
    \Require Atom positions of base ASU backbone $p_{bb}$, symmetry axes $l_1$, $l_2$, $l_5$, number of replicates $n$.
    \Statex 
    \Comment{$p_{bb} \in \mathbb{R}^{N\times 3\times 3}$, $l_1, l_2, l_5 \in \mathbb{R}^{2\times 3}$}
    \Ensure Atom positions of backbone $p_{all}$.
    \Statex
    \Comment{$p_{all} \in \mathbb{R}^{n\times N\times 3\times 3}$}

    \renewcommand{\algorithmicrequire}{\textcolor[RGB]{50,140,10}{def}}
    \Require ApplyTetrahedralSymmetry($p_{bb}$, $l_1$, $l_2$, $l_5$)

    \State $c_1 = \mathrm{mean}(p_{bb}[:,1,:])$
    \State $c_1,c_2,c_3 = \mathrm{ApplySymmetryOp}(c_1, l_1, 3)$

    \State $c_{123} = \mathrm{Concat}(c_1,c_2,c_3)$  
    
    \If{$l_2 \neq None$}
    \State $a_{12} = ((l_1[:3]-l_1[3:])^T(l_2[:3]-l_2[3:]))$
    \State $c_{123}, c_{456}, c_{789} = \mathrm{ApplySymmetryOp}(c_{123}, l_2, 3)$
    \State $p_1, p_2, p_3 = \mathrm{mean}(c_{123}),\mathrm{mean}(c_{456}),\mathrm{mean}(c_{789})$
    \State $d = ||p_1-p_2||$
    \State $u_n = l_1[:3]-l_1[3:]$
    \State $u_n = \frac{u_n}{||u_n||}$
    \State $c = p_1 + \sqrt{\frac{2}{3}}\cdot d \cdot u_n$
    \State $v_2, v_3 = p_2 - c, p_3 - c$
    \State $v_2, v_3 = v_2 - v_2^Tu_n\cdot u_n, v_3 - v_3^Tu_n\cdot u_n$
    \State $v_2, v_3 = \frac{v_2}{||v2||}\cdot \frac{\sqrt{3}}{3}\cdot d, \frac{v_3}{||v3||}\cdot \frac{\sqrt{3}}{3}\cdot d$
    \State $v_4 = -v_2-v_3$
    \State $v_4 = \frac{v_4}{||v4||}\cdot \frac{\sqrt{3}}{3}\cdot d$
    \State $p_4 = c + v_4$
    \State $T_{rot} = \mathrm{Frame3Points}(l_1[:3], c, p_4)$
    \State $v_{4n} = ||v_4||$
    \State $p_{2,local} = [0, -v_{4n}/2, \sqrt{3}/2\cdot v_{4n}]$
    \State $p_{3,local} = [0, -v_{4n}/2, -\sqrt{3}/2\cdot v_{4n}]$
    \State $p_2, p_3  = T_{rot}\circ p_{2,local},  T_{rot}\circ p_{3,local}$

    \Else
    
    \State $c_{123}, c_{456} = \mathrm{ApplySymmetryOp}(c_{123}, l_5, 2)$
    \State $p_1 = \mathrm{mean}(c_{123})$
    \State $p_2 = \mathrm{mean}(c_{456})$
    \State $d = ||p_1-p_2||$
    \State $u_n = l_1[:3]-l_1[3:]$
    \State $u_n = \frac{u_n}{||u_n||}$
    \State $c = p_1 + \sqrt{\frac{2}{3}}\cdot d \cdot u_n$
    \State $v_2 = p_2 - c$
    \State $v_2 = v_2 - v_2^Tu_n\cdot u_n$
    \State $T_{rot} = \mathrm{Frame3Points}(l_1[:3], c, p_4)$
    \State $v_{2n} = ||v_2||$
    \State $p_{3,local} = [0, -v_{2n}/2, \sqrt{3}/2\cdot v_{2n}]$
    \State $p_{4,local} = [0, -v_{24n}/2, -\sqrt{3}/2\cdot v_{2n}]$
    \State $p_3, p_4= T_{rot}\circ p_{3,local},  T_{rot}\circ p_{4,local}$
    \EndIf
    
    \State $v_0, v1, v2, v3, v4 = p_1, p_2, p_3, p_4$
    \State $m_1, m_2, m_3 = \frac{v_0+v_1}{2},\frac{v_0+v_2}{2},\frac{v_0+v_3}{2}$
    \State $m_{12}, m_{23}, m_{13} = \frac{v_1+v_2}{2}, \frac{v_2+v_3}{2}, \frac{v_1+v_3}{2}$
    \State $r_0 = [\frac{v_1+v_2+v_3}{3}, v_0]$
    \State $r_1, r_2, r_3 = [m_1, m_{23}], [m_2, m_{13}], [m_3, m_{12}]$
    \State $p_{bb} = \mathrm{ApplySymmetryOp}(p_{bb}, r_0, 3)$
    \State $p_{1} = \mathrm{ApplySymmetryOp}(p_{bb}, r_1, 2)$
    \State $p_{2} = \mathrm{ApplySymmetryOp}(p_{bb}, r_2, 2)$
    \State $p_{3} = \mathrm{ApplySymmetryOp}(p_{bb}, r_3, 2)$
    \State $p_{all} = \mathrm{concat}(p_{bb}, p_{1}, p_{2}, p_{3})$
    \State \Return $p_{all}$
    \end{algorithmic}
\end{algorithm*}

\begin{algorithm}[h]
\caption{Tetrahedral symmetry generator}
\label{alg_T_sym}
    \begin{algorithmic}[1]
    \renewcommand{\algorithmicrequire}{\textbf{Input:}}
    \renewcommand{\algorithmicensure}{\textbf{Output:}}
    \Require Backbone frames $\mathcal{T}$, Confidence score map $C$, Nearest position map $P^{nst}$, Chain id map $M$, Number of selection $K$, Number of replicates $n$.
    \Statex
    \Comment{$P^{nst} \in \mathbb{R}^{N\times N}$, $M \in \mathbb{R}^{7\times N\times N}$, $C \in \mathbb{R}^{7\times N\times N}$} 
    \Ensure Atom positions of backbone $p_{all}$.
    \Statex
    \Comment{$p_{all} \in \mathbb{R}^{n\times N\times 3\times 3}$}

    \renewcommand{\algorithmicrequire}{\textcolor[RGB]{50,140,10}{def}}
    \Require TetrahedralSymmetryGenerator($\mathcal{T}$, $P^{nst}$, $M$, $C$, $K$, $n$) 
    \State $p_{bb} := \text{backbone atom positions}$ \Comment{$p_{bb} \in \mathbb{R}^{N\times 3\times3 }$}    
    \State $p_0 := C^{\alpha} \text{ position}$ \Comment{$p_0 \in \mathbb{R}^{N\times 3 }$}
    \State $S = \mathrm{GetPointIndexSet}(M, C, K)$ \Comment{$S \in \mathbb{R}^{7\times K\times 2}$}
    \State $iso_2 = \sum S[1]$
    \State $hetero_2 = \sum S[5]$

    \If {$iso_2>hetero_2$} 
    \State $l_1 = \mathrm{SymmetryAxisIsologous}(\mathcal{T}, P^{nst}, S[0])$ 
    \State $l_2 = \mathrm{SymmetryAxisIsologous}(\mathcal{T}, P^{nst}, S[1])$ 
    \State $p_{1} = \mathrm{ApplySymmetryOp}(p_0, l_1, 2)$ 
    \State $p_{2} = \mathrm{ApplySymmetryOp}(p_0, l_2, 2)$ 
    \State $l_1, l_2 = \mathrm{SymmetryAxisIso2}(p_0, l_1, l_2, n)$ 
    \State $p_{all} = \mathrm{ApplyTetrahedralSymmetry}(p_{bb}, l_1, None, l_2)$
    
    \Else 
    \State $p_1 = \mathrm{ReconstructNeighbourChain}(\mathcal{T}, P^{nst}, S[3])$
    \State $p_2 = \mathrm{ReconstructNeighbourChain}(\mathcal{T}, P^{nst}, S[4])$
    \State $l_1 = \mathrm{SymmetryAxisHeterologous}(p_0, p_1, p_2)$ 
    \State $p_3 = \mathrm{ReconstructNeighbourChain}(\mathcal{T}, P^{nst}, S[5])$
    \State $p_4 = \mathrm{ReconstructNeighbourChain}(\mathcal{T}, P^{nst}, S[6])$
    \State $l_2 = \mathrm{SymmetryAxisHeterologous}(p_0, p_3, _p4)$ 
    \State $p_{all} = \mathrm{ApplyTetrahedralSymmetry}(p_{bb}, l_1, l_2, None)$
    \EndIf
    \State \Return $p_{all}$
    \end{algorithmic}
\end{algorithm}

\begin{algorithm}[h]
\caption{Generate the whole assembly by interfaces for octahedral symmetry}
\label{alg_apply_octahedral_symmetry}
    \begin{algorithmic}[1]
    \renewcommand{\algorithmicrequire}{\textbf{Input:}}
    \renewcommand{\algorithmicensure}{\textbf{Output:}}
    \Require Atom positions of base ASU backbone $p_{bb}$, symmetry axes $l_1$, $l_2$, number of replicates $n$.
    \Statex 
    \Comment{$p_{bb} \in \mathbb{R}^{N\times 3\times 3}$, $l_1, l_2 \in \mathbb{R}^{2\times 3}$}
    \Ensure Atom positions of backbone $p_{all}$.
    \Statex
    \Comment{$p_{all} \in \mathbb{R}^{n\times N\times 3\times 3}$}

    \renewcommand{\algorithmicrequire}{\textcolor[RGB]{50,140,10}{def}}
    \Require ApplyOctahedralSymmetry($p_bb$, $l_1$, $l_2$)
    
    \State $c = \mathrm{CrossPoint}(l_1, l_2)$
    \State $r_2 = (l_2[3:]-l_2[:3])\times (l_1[3:]-l_1[:3])$
    \State $r_2 = [c, r_2+c]$
    
    \State $c_{3} = \mathrm{ApplySymmetryOp}(p_{bb}, l_2, 3)$
    \State $c_{12} = \mathrm{ApplySymmetryOp}(c_{3}, l_1, 4)$
    \State $c_{24} = \mathrm{ApplySymmetryOp}(c_{12}, r_2, 2)$
    
    \State $p_{all} = c_{24}$
    
    \State \Return $p_{all}$
    \end{algorithmic}
\end{algorithm}

\begin{algorithm}[h]
\caption{Octahedral symmetry generator}
\label{alg_O_sym}
    \begin{algorithmic}[1]
    \renewcommand{\algorithmicrequire}{\textbf{Input:}}
    \renewcommand{\algorithmicensure}{\textbf{Output:}}
    \Require Backbone frames $\mathcal{T}$, Confidence score map $C$, Nearest position map $P^{nst}$, Chain id map $M$, Number of selection $K$, Number of replicates $n$.
    \Statex
    \Comment{$P^{nst} \in \mathbb{R}^{N\times N}$, $M \in \mathbb{R}^{7\times N\times N}$, $C \in \mathbb{R}^{7\times N\times N}$} 
    \Ensure Atom positions of backbone $p_{all}$.
    \Statex
    \Comment{$p_{all} \in \mathbb{R}^{n\times N\times 3\times 3}$}

    \renewcommand{\algorithmicrequire}{\textcolor[RGB]{50,140,10}{def}}
    \Require OctahedralSymmetryGenerator($\mathcal{T}$, $P^{nst}$, $M$, $C$, $K$, $n$) 
    \State $p_{bb} := \text{backbone atom positions}$ \Comment{$p_{bb} \in \mathbb{R}^{N\times 3\times3 }$}    
    \State $p_0 := C^{\alpha} \text{ position}$ \Comment{$p_0 \in \mathbb{R}^{N\times 3 }$}
    \State $S = \mathrm{GetPointIndexSet}(M, C, K)$ \Comment{$S \in \mathbb{R}^{7\times K\times 2}$}

    \State $p_1 = \mathrm{ReconstructNeighbourChain}(\mathcal{T}, P^{nst}, S[3])$
    \State $p_2 = \mathrm{ReconstructNeighbourChain}(\mathcal{T}, P^{nst}, S[4])$
    \State $l_1 = \mathrm{SymmetryAxisHeterologous}(p_0, p_1, p_2)$ 
    \State $p_3 = \mathrm{ReconstructNeighbourChain}(\mathcal{T}, P^{nst}, S[5])$
    \State $p_4 = \mathrm{ReconstructNeighbourChain}(\mathcal{T}, P^{nst}, S[6])$
    \State $l_2 = \mathrm{SymmetryAxisHeterologous}(p_0, p_3, p_4)$ 
    \State $p_{all} = \mathrm{ApplyOctahedralSymmetry}(p_{bb}, l_1, l_2)$
    
    \State \Return $p_{all}$
    \end{algorithmic}
\end{algorithm}

\begin{algorithm}[h]
\caption{Generate the whole assembly by interfaces for icosahedral symmetry}
\label{alg_apply_icosahedral_symmetry}
    \begin{algorithmic}[1]
    \renewcommand{\algorithmicrequire}{\textbf{Input:}}
    \renewcommand{\algorithmicensure}{\textbf{Output:}}
    \Require Atom positions of base ASU backbone $p_{bb}$, symmetry axes $l_1$, $l_2$,  $l_5$, number of replicates $n$.
    \Statex 
    \Comment{$p_{bb} \in \mathbb{R}^{N\times 3\times 3}$, $l_1, l_2, l_5 \in \mathbb{R}^{2\times 3}$}
    \Ensure Atom positions of backbone $p_{all}$.
    \Statex
    \Comment{$p_{all} \in \mathbb{R}^{n\times N\times 3\times 3}$}

    \renewcommand{\algorithmicrequire}{\textcolor[RGB]{50,140,10}{def}}
    \Require ApplyIcosahedralSymmetry($p_{bb}$, $l_5$, $l_3$, $l_2$)
    
    \State $r_{2s} = l_2[:3]$ 
    \State $r_{2e} = l_2[3:]$ 
    \State $r12_{2s} = \mathrm{ApplySymmetryOp}(r_{2s}, l_3, 3)$     
    \State $r12_{2e} = \mathrm{ApplySymmetryOp}(r_{2e}, l_3, 3)$
    \State $r2_1 = [r12_{2s}[1], r12_{2e}[1]]$ 
    \State $r2_2 = [r12_{2s}[2], r12_{2e}[2]]$
    \State $\theta_1 = \mathrm{Angle}(r2_1, l_5)$ 
    \State $\theta_2 = \mathrm{Angle}(r2_2, l_5)$

    \State \If{$\theta_1 \geq \theta_2$}
    \State $r_v=r2_1$
    \State \Else
    \State $r_v=r2_2$
    \State \EndIf
    
    \State $c = \mathrm{CrossPoint}(l_5, l_3)$
    \State $r_2 = (l_5[3:]-l_5[:3])\times (l_3[3:]-l_3[:3])$
    \State $r_2 = [c, r_2+c]$
    
    \State $c_{3} = \mathrm{ApplySymmetryOp}(p_{bb}, l_3, 3)$
    \State $c_{6} = \mathrm{ApplySymmetryOp}(c_{3}, r_v, 2)$
    \State $c_{30} = \mathrm{ApplySymmetryOp}(c_{6}, l_5, 5)$
    \State $c_{60} = \mathrm{ApplySymmetryOp}(c_{6}, r_2, 2)$
    
    \State $p_{all} = c_{60}$
    
    \State \Return $p_{all}$
    \end{algorithmic}
\end{algorithm}

\begin{algorithm}[h]
\caption{Icosahedral symmetry generator}
\label{alg_I_sym}
    \begin{algorithmic}[1]
    \renewcommand{\algorithmicrequire}{\textbf{Input:}}
    \renewcommand{\algorithmicensure}{\textbf{Output:}}
    \Require Backbone frames $\mathcal{T}$, Confidence score map $C$, Nearest position map $P^{nst}$, Chain id map $M$, Number of selection $K$, Number of replicates $n$.
    \Statex
    \Comment{$P^{nst} \in \mathbb{R}^{N\times N}$, $M \in \mathbb{R}^{7\times N\times N}$, $C \in \mathbb{R}^{7\times N\times N}$} 
    \Ensure Atom positions of backbone $p_{all}$.
    \Statex
    \Comment{$p_{all} \in \mathbb{R}^{n\times N\times 3\times 3}$}

    \renewcommand{\algorithmicrequire}{\textcolor[RGB]{50,140,10}{def}}
    \Require IcosahedralSymmetryGenerator($\mathcal{T}$, $P^{nst}$, $M$, $C$, $K$, $n$) 
    \State $p_{bb} := \text{backbone atom positions}$ \Comment{$p_{bb} \in \mathbb{R}^{N\times 3\times3 }$}   
    \State $p_0 := C^{\alpha} \text{ position}$ \Comment{$p_0 \in \mathbb{R}^{N\times 3 }$}
    \State $S = \mathrm{GetPointIndexSet}(M, C, K)$ \Comment{$S \in \mathbb{R}^{7\times K\times 2}$}

    \State $p_1 = \mathrm{ReconstructNeighbourChain}(\mathcal{T}, P^{nst}, S[3])$
    \State $p_2 = \mathrm{ReconstructNeighbourChain}(\mathcal{T}, P^{nst}, S[4])$
    \State $l_1 = \mathrm{SymmetryAxisHeterologous}(p_0, p_1, p_2)$ 
    \State $p_3 = \mathrm{ReconstructNeighbourChain}(\mathcal{T}, P^{nst}, S[5])$
    \State $p_4 = \mathrm{ReconstructNeighbourChain}(\mathcal{T}, P^{nst}, S[6])$
    \State $l_2 = \mathrm{SymmetryAxisHeterologous}(p_0, p_3, p_4)$
    \State $l_3 = \mathrm{SymmetryAxisIsologous}(\mathcal{T}, P^{nst}, S[0])$ 
    \State $p_{all} = \mathrm{ApplyIcosahedralSymmetry}(p_{bb}, l_1, l_2, l_3)$
    
    \State \Return $p_{all}$
    \end{algorithmic}
\end{algorithm}

\end{document}